\documentclass[fontsize=12pt,a4paper,headings=normal,
twoside=false,leqno,parskip=half-,abstract=true]{scrartcl}
\usepackage[english]{babel}
\usepackage{mdframed}
\usepackage[utf8]{inputenc}
\usepackage{hyperref}
\hypersetup{
 pdftitle={},
 pdfauthor={},
 colorlinks=true,
 linkcolor=blue,
 citecolor=blue,
 filecolor=blue,
 urlcolor=blue}

\usepackage[pagewise]{lineno}
\usepackage[version=4]{mhchem}
\usepackage{mathtools} 
\usepackage[format=plain,labelfont=bf,font=small]{caption}
\usepackage{subfigure}
\usepackage{xcolor}
\usepackage[arrow, matrix, curve]{xy}
\usepackage{tikz}
\usetikzlibrary{arrows.meta, decorations.pathmorphing}
\usepackage{float}
\usepackage{orcidlink}

\usepackage{academicons}
\definecolor{orcidlogocol}{HTML}{A6CE39}
\usepackage{tikz-cd}
\usepackage{caption}
\usepackage{blkarray}
\usepackage{amsmath,amsthm}
\usepackage{amssymb} 
\usepackage[normalem]{ulem}
\usepackage{enumitem}
\usepackage[normalem]{ulem}
\usepackage[version=4]{mhchem}

\usepackage{algorithm}%
\usepackage{algorithmicx}%
\usepackage{algpseudocode}%
\usepackage{verbatim}
\newtheorem{theorem}{Theorem}[section]
\newtheorem{lemma}[theorem]{Lemma}

\newtheorem{prop}[theorem]{Proposition}

\theoremstyle{definition}
\newtheorem{definition}[theorem]{Definition}

\theoremstyle{remark}

\numberwithin{equation}{section}

\DeclareMathAlphabet{\mathpzc}{OT1}{pzc}{m}{it}

\newcommand{\be}{\begin{equation}}
\newcommand{\ee}{\end{equation}}
\newcommand{\bea}{\begin{eqnarray}}
\newcommand{\eea}{\end{eqnarray}}

\title{\textsc{BiRNe}:\\
Symbolic bifurcation analysis of reaction networks with Python}
\author{Richard Golnik\thanks{Universität Leipzig, Germany \texttt{richard@bioinf.uni-leipzig.de}}\;, Thomas Gatter\thanks{Universität Leipzig, Germany \texttt{thomas@bioinf.uni-leipzig.de}}\;, Peter F. Stadler\thanks{Universität Leipzig, Germany \texttt{studla@bioinf.uni-leipzig.de}}\;, \\ Nicola Vassena \thanks{Universität Leipzig, Germany, \texttt{nicola.vassena@uni-leipzig.de}}}

\date{\today}

\begin{document}
\maketitle

\begin{abstract}
Computer algebra methods for analyzing reaction networks often rely on the assumption of mass-action kinetics, which transform the governing ODEs into polynomial systems amenable to techniques such as Gröbner basis computation and related algebraic tools. However, these methods face significant computational complexity, limiting their applicability to relatively small networks involving only a handful of species.
In contrast, building on recent theoretical advances, we introduce here  \textsc{BiRNe} (BIfurcations in Reaction NEtworks) Python module, which relies on a symbolic approach designed to detect bifurcations in larger reaction networks (up to 10-20 species, depending on the network's connectivity) equipped with parameter-rich kinetics. This class includes enzymatic kinetics such as Michaelis--Menten, ligand-binding kinetics like Hill functions, and generalized mass-action kinetics. For a given network, the current algorithm identifies all minimal autocatalytic subnetworks and fully characterizes the presence of bifurcations associated with zero eigenvalues, thus determining whether the network admits multistationarity. It also detects oscillatory bifurcations arising from positive-feedback structures, capturing a significant class of possible oscillations.
\end{abstract}

\section{Introduction}

Reaction networks naturally arise as modeling structures in various fields, including biology, chemistry, ecology, economics, and epidemiology. A primary difficulty in analyzing such systems is the widespread lack of knowledge about the governing mathematical laws and, even more so, about the relevant quantities involved. For this reason, it is customary to consider classes of systems that depend on many parameters. Since realistic reaction networks are typically high-dimensional, the number of parameters is also large, and the computational complexity of even simple tasks quickly becomes intractable.

Historically, a large body of work has focused on mass-action kinetics, which models interactions as monomials in the concentrations of the reacting species. This assumption is particularly meaningful when considering elementary reactions in well-mixed reactors or, for example, in epidemiological models describing transition rates between compartmental classes. However, it may become less appropriate in other contexts, such as biochemical or cellular systems. Under mass-action kinetics, analyzing the fixed points of the concentration system amounts to solving for the roots of a system of polynomial equations, for which many tools from algebraic geometry are available. The main difficulty, however, is that analyzing fixed-point stability requires first obtaining a parametrization of the fixed-point variety, on which all subsequent analysis must then be performed. This unavoidable step comes with a great computational cost even for simple tasks, such as assessing the stability of fixed points, let alone more delicate bifurcation problems. For an illustration of the typical difficulties involved, see, among many others, Hernandez et al~\cite{hernandez23} for parametrization of fixed-points, Conradi et al.~\cite{Conradi2007}, Pérez Millán \& Dickenstein~\cite{MESSI:18}, and Röst \& Sadeghimanesh \cite{rost21} for zero-eigenvalue bifurcations, and Gatermann et al.~\cite{Gat2005} and Banaji \& Borosz~\cite{BaBo23} for purely imaginary-eigenvalue bifurcations.

In this contribution, we present an alternative computational tool that relies solely on the linearization of the system, without requiring explicit computation of fixed points. In particular, this method applies to any system for which the characteristic polynomial of the Jacobian can be obtained in symbolic form: typically covering networks of up to around 15 species. The central assumption is that the kinetics are \emph{not} in mass-action form, but instead belong to a broad class of so-called \emph{parameter-rich kinetics} \cite{VasStad23}, which includes widely used schemes such as Michaelis–Menten, Hill, and generalized mass-action kinetics.

This paper is organized as follows. Section~\ref{sec:preliminaries} presents the standard literature ingredients concisely and refers to the appropriate sources. Section~\ref{sec:birne} introduces our newly developed Python module. An example of its applicability is presented in Section~\ref{sec:example}. The discussion Section~\ref{sec:outlook} closes the paper.

\textbf{Acknowledgment.} We thank AmirHosein Sadeghimanesh for his continued and successful organization of the session on Computer Algebra Applications in the Life Sciences (CASinLife) at the ACA conferences, where this work was presented. This work has been supported by the MATOMIC consortium,
  funded by the Novo Nordisk Foundation, grant NNF21OC0066551.

\section{Preliminary theory}\label{sec:preliminaries}

The content of this section has been presented in more detail in the publications \cite{VasHunt,VasStad23,blokhuis_stoichiometric_2025}. 

\subsection{Reaction networks and symbolic Jacobians} 
We consider reaction networks $\mathbf{\Gamma}$ as pairs of sets $\mathbf{\Gamma}=(M,E)$, where $M$ is the set of species $X_m$ and $E$ is the set of reactions $j$. Any reaction $j$ is an ordered association between nonnegative linear combinations of the species:
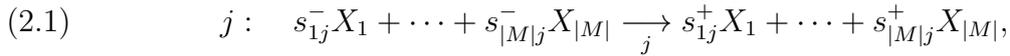
\begin{equation} \label{reactionj}
 j:\quad s^-_{1j}X_1+\dots+s^-_{|M|j}X_{|M|} \underset{j}{\longrightarrow} {s}^+_{1j}X_1+\dots+{s}^+_{|M|j}X_{|M|},
\end{equation}
where the nonnegative coefficients $s^-_{mj}, s^+_{mj} \geq 0$ are the \emph{stoichiometric coefficients}. Species $X_m$ appearing on the left-hand side (resp. right-hand side) of \eqref{reactionj} with nonzero stoichiometric coefficients $s^-_{mj}>0$ (resp. $s^{+}_{mj}>0$) are called \emph{reactants} (resp. \emph{products}). The stoichiometric coefficients of the reactant and product species are collected in two $|M|\times |E|$ matrices. The \emph{reactant matrix} ${S}^{-}$ is defined by
\begin{equation}\label{eq:reactantmatrix}
{S}^-_{mj} \coloneqq s^-_{mj},
\end{equation}
and the \emph{product matrix} $S^+$ by
\[\label{eq:productmatrix}
S^+_{mj} \coloneqq s^+_{mj}.
\]
The difference of the two matrices gives rise to the \emph{stoichiometric matrix} $S:=S^+ -S^-$ with entries
\begin{equation}\label{eq:stoichmatrix}
   S_{mj} \coloneqq s_{mj}^+ - s_{mj}^-.
\end{equation}

Let $x(t)\in\mathbb{R}^{|M|}_{\ge 0}$ denote the time evolution of the nonnegative vector of species concentrations in a well-mixed, spatially homogeneous reactor. Its dynamics follow the system of Ordinary Differential Equations (ODEs)
\begin{equation}\label{eq:maineq}
    \dot{x}=f(x)=S\mathbf{r}(x),
\end{equation}
where $S$ is the stoichiometric matrix defined in \eqref{eq:stoichmatrix} and $\mathbf{r}(x) \in \mathbb{R}^{|E|}_{\ge 0}$ is the vector of \emph{reaction rates}. In applications, the precise functional form of $r_j(x)$ is typically unknown. For this reason, the literature usually resorts to parametric families $\mathbf{r}(x,p)$, where $p$ denotes positive parameters. Widely used reaction schemes in (bio)chemistry, ecology, and epidemiology include classic and generalized mass-action kinetics \cite{MA64,Muller:12}, Michaelis--Menten kinetics \cite{MM13}, and Hill kinetics \cite{Hill10}. In ecology, these correspond respectively to Holling Functional Responses of type I, II, and III \cite{Holling65}. In line with this tradition, we restrict attention to reaction rates that are \emph{monotone chemical functions} in the following sense.

\begin{definition}[Monotone chemical function]\label{def:monochemfunction}
  A function $r_j$ is a monotone chemical function if:
  \begin{enumerate}
  \item[i.] $r_j(x) \geq 0$ for all $x\in\mathbb{R}^{|M|}_{\ge0}$;
  \item[ii.] $r_j(x)>0$ implies $x_m>0$ for all species $X_m$ with $s^-_{mj}>0$;
  \item[iii.] $s^-_{mj}=0$ implies $\partial r_j/\partial x_m \equiv 0$;
  \item[iv.] for $x>0$ and $s^-_{mj}>0$, it holds that $\partial r_j/\partial x_m>0$.
  \end{enumerate}
\end{definition}

A parametric vector $\mathbf{r}(x,p)$ of reaction rates for $\mathbf{\Gamma}$ such that $r_j(x,p)$ is a monotone chemical function for all $j$ and for any choice of admissible parameters is called a \emph{monotone kinetic model} for the network $\mathbf{\Gamma}$.

Let now $\bar{x}$ be a fixed point (steady state) of \eqref{eq:maineq}, i.e.
\begin{equation}
    0=f(\bar{x})=S\mathbf{r}(\bar{x}).
\end{equation}
Its stability and possible bifurcations can be addressed at a linear approximation via the Jacobian matrix $f_x$ of partial derivatives evaluated at $\bar{x}$:
\begin{equation}
    f_x=\dfrac{\partial f}{\partial x}\bigg|_{x=\bar{x}}=S\dfrac{\partial \mathbf{r}}{\partial x}\bigg|_{x=\bar{x}},
\end{equation}
where $\dfrac{\partial \mathbf{r}}{\partial x}$ is the $|E|\times|M|$ \emph{reactivity matrix}. Assumptions (iii) and (iv) in Def.~\ref{def:monochemfunction} imply that, at positive concentrations $x>0$, the evaluated reactivity matrix is nonnegative:
\begin{equation}\label{eq:reactivitymatrix}
  \bigg(  \dfrac{\partial \mathbf{r}}{\partial x}\bigg)_{jm}=\dfrac{\partial r_j}{\partial x_m}=\begin{cases}
      >0 &\text{if $X_m$ is reactant to $j$, i.e. $s^-_{mj}>0$;}\\
      =0 &\text{otherwise.}
  \end{cases}
\end{equation}

Inspired by \eqref{eq:reactivitymatrix}, we define the $|E|\times |M|$ \emph{symbolic reactivity matrix} $R$ by
\begin{equation}\label{eq:symreactivitymatrix}
    R_{jm}=\begin{cases}
        r'_{jm}>0 &\text{if $X_m$ is reactant to $j$, i.e. $s^-_{mj}>0$;}\\
      =0 &\text{otherwise,}
    \end{cases}
\end{equation}
where a positive symbol $r'_{jm}$ represents a positive parameter. The reactivity matrix is thus a symbolic version of the transpose of the reactant matrix \eqref{eq:reactantmatrix}, as stated below.

\begin{prop}
For the symbolic reactivity matrix $R$, it holds that
\[
R_{jm}> 0 \quad\Leftrightarrow\quad S^-_{mj}>0.
\]
\end{prop}
\proof
Directly by Eq.~\eqref{eq:symreactivitymatrix}.
\endproof

Accordingly, we define the \emph{symbolic Jacobian matrix} $G$ as
\begin{equation}\label{eq:symbolicjacobian}
    G:=SR.
\end{equation}

In the theory presented here, we use the symbolic Jacobian to investigate two types of results: \emph{exclusion results} and \emph{existence results}.
\begin{enumerate}
    \item \textbf{Exclusion results:} hold irrespective of the choice of symbols in the symbolic reactivity matrix $R$;
    \item \textbf{Existence results:} hold for specific choices of the symbols in $R$.
\end{enumerate}
Exclusion results apply to any network endowed with a monotone kinetic model. Existence results, however, depend on the choice of kinetic model and require thus further specification.

To clarify, assume that for a certain network $\mathbf{\Gamma}$, the symbolic Jacobian $G=SR$ is invertible for all admissible $R$. Since any fixed-point Jacobian corresponds to a specific instantiation of $R$, we can \emph{exclude} the possibility that any fixed-point in any monotone kinetic model has a singular Jacobian.  
Conversely, assume that $G=SR$ can be made singular for specific choices of $R$. The fact that such an $R$ corresponds to an actual fixed-point reactivity matrix is not guaranteed \emph{a priori}. First, the existence of any fixed-point requires a positive right kernel vector $v\in\mathbb{R}^{|E|}_{>0}$ of the stoichiometric matrix $S$:
\begin{equation}\label{eq:consistent}
    Sv=0.
\end{equation}
Networks satisfying \eqref{eq:consistent} are called \emph{consistent} \cite{Ang07}, and existence results based on fixed-point Jacobians require consistency: this is the case of fixed-point bifurcation results. Second, even for consistent networks, mapping the symbols $r_{jm}$ in $R$ to explicit parameters $p$ in $\mathbf{r}(x,p)$ is not trivial. The notion of \emph{parameter-rich kinetics} \cite{VasStad23} identifies those monotone kinetic models for which this mapping is always possible.

\begin{definition}[Parameter-rich kinetics]\label{def:prich}
  A monotone kinetic rate model $\mathbf{r}(x,p)$ is called \emph{parameter-rich} if, for
  every positive fixed-point $\bar{x}>0$ and every choice of an $|E|\times
  |M|$ symbolic reactivity matrix $\bar{R}$, there exists a choice of parameters $\bar{p} = p(\bar{x},
  \bar{R})$ such that $\partial r_j/\partial x_m (\bar{x}, \bar{p}) = \bar{R}_{jm}.$
\end{definition}

\begin{mdframed}
\begin{minipage}[t]{0.48\textwidth}
\textbf{EXCLUSION RESULTS.}

\vspace{0.4cm}

\begin{equation*}
\underset{1}\longrightarrow \quad X_1 \quad \underset{2}\longrightarrow \quad X_2 \quad \underset{3}\longrightarrow
\end{equation*}

\vspace{0.4cm}

Equations:
\begin{equation*}
\begin{cases}
\dot{x}_1=K_1-r_2(x_1),\\
\dot{x}_2=r_2(x_1)-r_3(x_2).
\end{cases}
\end{equation*}
Symbolic Jacobian:
\begin{equation*}
    \begin{pmatrix}
    -r'_{1x_1}  & 0\\
    r'_{1x_1} & -r'_{2x_2}
    \end{pmatrix}.
\end{equation*}
The eigenvalues are strictly negative irrespective of any parameter and concentration values. Fixed-point bifurcations are thus excluded for any choice of monotone kinetic model, and any fixed-point is locally stable.\\

\vspace{0.5cm}

\emph{Exclusion results are always valid also for mass-action kinetics.}
\end{minipage}\hspace{1em}%
\vrule width 0.5pt
\hspace{1em}%
\begin{minipage}[t]{0.48\textwidth}
\textbf{EXISTENCE RESULTS.}
\begin{equation*}
\begin{cases}
    X_1 \quad\underset{1}\longrightarrow\quad &X_2 \quad \underset{2}\longrightarrow \quad 2X_1,\\
    \textcolor{white}{X_1}\quad\underset{3}\longrightarrow \quad &X_1\quad \underset{4}\longrightarrow 
\end{cases}
\end{equation*}
Equations:
\begin{equation*}
\begin{cases}
\dot{x}_1=K_3-r_1(x_1)+2r_2(x_2)-r_4(x_1),\\
\dot{x}_2=r_1(x_1)-r_2(x_2).
\end{cases}
\end{equation*}

Symbolic Jacobian:
\begin{equation*}
    \begin{pmatrix}
    -r'_{1x_1} - r'_{4x_1} & 2r'_{2x_2}\\
    r'_{1x_1} & -r'_{2x_2}
    \end{pmatrix}.
\end{equation*}
As $\operatorname{det}=r'_{2x_2}(r'_{4x_1}-r'_{1x_1})$, a zero-eigenvalue bifurcation requires $r'_{4x_1}=r'_{1x_1}$. Definition~\ref{def:prich} guarantees that this condition is always realizable at a fixed-point for parameter-rich kinetics. See \cite{V23} for an explicit realization of a saddle-node bifurcation for Michaelis--Menten kinetics.\\

\emph{Existence results are not always valid for mass-action kinetics.}
\end{minipage}
\end{mdframed}

\subsection{Child-Selections and bifurcations}

The previous subsection introduced the concepts of the \emph{symbolic Jacobian} and \emph{parameter--rich kinetics}. We now present the main tool for studying the spectra of symbolic Jacobians and possible bifurcations: \emph{Child-Selections}.

\begin{definition}[\textbf{Child Selection}]
A \emph{$k$-Child-Selection triple} (or \emph{$k$-CS}) is a triple
$\pmb{\kappa}=(\kappa, E_\kappa, J)$ such that
$|\kappa|=|E_\kappa|=k$, $\kappa\subseteq M$, $E_\kappa\subseteq E$, and
$J:\kappa\to E_\kappa$ is a bijection satisfying
$s_{mJ(X_m)}^->0$ for all $X_m\in \kappa$.  
The map $J$ is called the \emph{Child-Selection bijection}.
\end{definition}

For each $k$-CS $\pmb{\kappa}$, we define the associated
$k\times k$ \emph{Child-Selection matrix} (or \emph{CS-matrix}) as
\begin{equation}
  S[\pmb{\kappa}]_{ml}\coloneqq s^+_{mJ(l)}-s^-_{mJ(l)}.
\end{equation}
A CS-matrix is thus a square submatrix of the stoichiometric matrix $S$, up to a reshuffling of its columns according to the CS-bijection $J$. Rows follow the ordering of the species in $M$, while columns follow the order induced by $J$.

To analyze the spectrum of the symbolic Jacobian $G$, we consider its characteristic polynomial
\begin{equation}\label{eq:charpoly}
    g(\lambda)=\det(\lambda\operatorname{Id} -G)=\sum_{k=1}^{|M|}a_k (-1)^k\lambda^{|M|-k},
\end{equation}
and employ a central lemma from \cite{VasHunt, VasStad23}, which expands each coefficient $a_k$ in terms of Child-Selections.

\begin{lemma}\label{lem:CSexp}
Let $g(\lambda)$ be the characteristic polynomial of the symbolic Jacobian $G$ of a network $\mathbf{\Gamma}$.  
For each coefficient $a_k$ in \eqref{eq:charpoly}, it holds that
\begin{equation}
a_k=\sum_{\pmb{\kappa}}\operatorname{det}S[\pmb{\kappa}]\prod_{X_m\in\kappa} R_{J(X_m)m},
\end{equation}
where the sum runs over all $k$-CS triples $\pmb{\kappa}$.
\end{lemma}

The proof is based on the Cauchy--Binet formula and for this reason each summand $\operatorname{det}S[\pmb{\kappa}]\prod_{X_m\in\kappa}$ is called a \emph{Cauchy--Binet summand} (or CB summand). Lemma~\ref{lem:CSexp} provides the foundation for deriving spectral properties of the Jacobian using Child-Selections alone. By selecting appropriate subsets $\pmb{\kappa}$, one can approximate or reconstruct the characteristic polynomial of $G$ through the corresponding CS-matrices. In the following, we list the main implications of this approach, which can be automatically explored by our Python module.

\paragraph{Nondegenerate networks.}
Most stability and bifurcation results consider only \emph{nondegenerate networks}, defined as follows.  
Consider an initial condition $x_0 \in \mathbb{R}^{|M|}_{>0}$ for \eqref{eq:maineq}. The sets
\begin{equation}
x_0 + \operatorname{Im} S,
\end{equation}
which are invariant under the flow, are called \emph{stoichiometric compatibility classes}.  

Left-kernel vectors $w$ of the stoichiometric matrix $S$ correspond to \emph{linear conservation laws}, since
\begin{equation}
\dfrac{d (w x(t))}{dt} = w\dot{x} = wS\mathbf{r}(x) = 0.
\end{equation}
Arbitrary but fixed values for these conservation laws, i.e. $C_i \in \mathbb{R}$ such that $w_i x(t) = C_i$, can be used to reduce the system to a specific stoichiometric compatibility class via a standard reduction procedure.  

A network is \emph{nondegenerate} if the dimension of its associated dynamical system equals the number of species $|M|$ minus the dimension $n$ of the left kernel of $S$, i.e. the number of independent linear conservation laws. Equivalently:

\begin{definition}[Nondegenerate networks]\label{def:nondegnet}
Let $\mathbf{\Gamma}$ be a reaction network with $|M|\times|E|$ stoichiometric matrix $S$, and let $n \ge 0$ denote the number of linearly independent conservation laws, i.e. 
$n = \operatorname{dim}\operatorname{ker}S^T$.  
Then $\mathbf{\Gamma}$ is \emph{nondegenerate} if there exists a choice of the reactivity matrix $R$ such that the symbolic Jacobian $G = SR$ possesses a nonzero $(|M|-n)$-principal minor.
\end{definition}

In terms of Child-Selections, the following characterization holds.
\begin{prop}[\cite{blokhuis_stoichiometric_2025}]\label{prop:nondegnet}
In the setting of Def.~\ref{def:nondegnet}, a network $\mathbf{\Gamma}$ is nondegenerate if and only if there exists an invertible $(|M|-n)\times(|M|-n)$ CS-matrix.
\end{prop}

Similarly, the spectrum of specific CS-matrices provides information about the possible spectral configurations of the symbolic Jacobian, as clarified below.

\paragraph{Instability and stability.}

Recall that a matrix $A$ is \emph{Hurwitz-stable} if all its eigenvalues have negative real part, and \emph{Hurwitz-unstable} if at least one eigenvalue has positive real part. In dynamical systems theory \cite{Hsubook}, the Hurwitz-(in)stability of the Jacobian matrix sufficiently determines the local (in)stability of the corresponding fixed point.  
In our parameter-rich framework, this motivates the following definitions.

\begin{definition}[Network admits stability]\label{def:admitsstab}
A nondegenerate network $\mathbf{\Gamma}$ \emph{admits stability} if there exists a choice of the symbolic reactivity matrix $\bar{R}$ such that the symbolic Jacobian $S\bar{R}$ has $|M|-n$ eigenvalues with negative real part, counted with algebraic multiplicity. Here, $n$ again denotes the number of independent conservation laws.
\end{definition}

\begin{definition}[Network admits instability]
A network $\mathbf{\Gamma}$ \emph{admits instability} if there exists a choice of the symbolic reactivity matrix $\bar{R}$ such that the corresponding symbolic Jacobian $S\bar{R}$ is Hurwitz-unstable.
\end{definition}

Definition~\ref{def:admitsstab} can be interpreted as stating that there exists a choice of the symbolic reactivity matrix such that the symbolic Jacobian, when restricted to a stoichiometric compatibility class, is Hurwitz-stable.  
Consequently, if a consistent and nondegenerate network admits stability, then there exists a parameter choice for which \eqref{eq:maineq} possesses a stable fixed point. Conversely, if a consistent network endowed with parameter-rich kinetics admits instability, then there exists a parameter choice for which system \eqref{eq:maineq} possesses an unstable fixed point.   
Note that concluding directly on stability requires nondegeneracy, unlike the unstable case.

In terms of CS-matrices, the following propositions provide sufficient conditions.

\begin{prop}[\cite{VasHunt}]\label{prop:CSstable}
For a nondegenerate network, if there exists a $(|M|-n)\times(|M|-n)$ CS-matrix $S[\pmb{\kappa}]$ that is Hurwitz-stable, then the network admits stability.
\end{prop}

\begin{prop}[\cite{VasStad23}]\label{prop:CSunstable}
If there exists a CS-matrix $S[\pmb{\kappa}]$ that is Hurwitz-unstable, then the network admits instability.
\end{prop}

Both Prop.~\ref{prop:CSstable} and Prop.~\ref{prop:CSunstable} can be generalized by allowing the existence of a positive diagonal matrix $D$ such that $S[\pmb{\kappa}]D$ is Hurwitz-(un)stable. The current version of our Python module, however, does not yet implement this generalization, and therefore we present the propositions in this restricted form.

\paragraph{Cores.} Let $\pmb{\kappa}=(\kappa, E_{\kappa}, J)$ be a $k$-CS.  
Child-Selections can be concatenated in the sense that any restriction $\pmb{\kappa'}$ of $\pmb{\kappa}$, to subsets $\kappa' \subset \kappa$ and $E_{\kappa'} = J(\kappa') \subset E_{\kappa}$, is itself a $k'$-CS $\pmb{\kappa'}=(\kappa', E_{\kappa'}, J)$.  
From the perspective of CS-matrices, the CS-matrix associated with $\pmb{\kappa'}$ appears as a principal submatrix of the CS-matrix associated with $\pmb{\kappa}$.

Therefore, it is natural to consider \emph{minimal} CS-matrices $S[\pmb{\kappa}]$ with respect to a matrix property $\mathbb{P}$, meaning that $S[\pmb{\kappa}]$ satisfies $\mathbb{P}$ while no proper principal submatrix does. For examples, in \cite{VasStad23} \emph{unstable cores} are defined as Hurwitz-unstable CS-matrices such with no Hurwitz-unstable principal submatrix. For the purposes of this study, the main property of interest concerns specifically the sign of the determinant of unstable cores.

\begin{definition}[Unstable-positive feedback, \cite{VasStad23}]\label{def:unstable-positive}
A $k\times k$ CS-matrix $S[\pmb{\kappa}]$ is an \emph{unstable-positive feedback} if
\begin{equation}
\operatorname{sign}\det S[\pmb{\kappa}] = (-1)^{k-1},
\end{equation}
and no $k'<k$ principal submatrix $S[\pmb{\kappa}']$ satisfies
\begin{equation}
\operatorname{sign}\det S[\pmb{\kappa}'] = (-1)^{k'-1}.
\end{equation}
\end{definition}

Unstable-positive feedbacks always possess a single real positive eigenvalue \cite{VasStad23}.  
In \cite{VasStad23}, the definition was slightly more restrictive, requiring additionally that no principal submatrix of $S[\pmb{\kappa}]$ be Hurwitz-unstable.  
For the purposes of the present analysis, Def.~\ref{def:unstable-positive} is the most appropriate.

For completeness, we also recall the definition of \emph{unstable-negative feedback}, although the current version of our Python module \textsc{BiRNe} primarily focuses on unstable-positive feedbacks.

\begin{definition}[Unstable-negative feedback, \cite{VasStad23}]\label{def:unstable-negative}
A $k\times k$ CS-matrix $S[\pmb{\kappa}]$ is an \emph{unstable-negative feedback} if it is Hurwitz-unstable and
\begin{equation}
\operatorname{sign}\det S[\pmb{\kappa}] = (-1)^{k},
\end{equation}
and no $k'<k$ principal submatrix $S[\pmb{\kappa}']$ is Hurwitz-unstable.
\end{definition}

Unstable-negative feedbacks always have one pair of complex-conjugate eigenvalues with positive real part and no real positive eigenvalue \cite{VasStad23}.

\paragraph{Autocatalysis.}

Autocatalysis is a fundamental concept in chemistry. Blokhuis et al. \cite{blokhuis20} provided a characterization of autocatalysis from a stoichiometric perspective, which has been then characterized in \cite{VasStad23} in terms of CS-matrices. We recall the relevant definition.
\begin{definition}[Stoichiometric Autocatalysis.]
A network $\mathbf{\Gamma}$ with stoichiometric matrix $S$ is \emph{autocatalytic} if there exists a $|M'|\times |E'|$ submatrix $S_A$ of $S$ such that 
\begin{enumerate}
    \item for any reaction $j$ appearing as a column of $S_A$, we have $m_1$ and $m_2$, not necessarily distinct, with $s^-_{m_1j}s^+_{m_2j}\neq 0$.
    \item there exists a positive vector $v\in\mathbb{R}^{|E'|}_{>0}$ such that
    \begin{equation}
    Sv>0.
\end{equation}
\end{enumerate}
\end{definition}
We also recall that a square matrix $A$ is called \emph{Metzler} if all its nondiagonal entries are nonnegative, i.e. $A_{ij}\ge0$, for all $i,j$ \cite{Bullo20}. We have the following result, which characterizes autocatalysis in terms of CS-matrices. 
\begin{theorem}[\cite{VasStad23}]\label{thm:autocat}
A network is autocatalytic if and only if there exists a CS-matrix $S[\pmb{\kappa}]$ which is an unstable-positive feedback and a Metzler matrix.
\end{theorem}

\paragraph{Real-zero eigenvalue: Multistationarity.}

Multistationarity refers to the coexistence of two or more fixed points under identical parameter values, and has been identified as a central mechanism underlying, for example, cell differentiation \cite{ThomKauf01}. A network endowed with a parametric kinetic model is said to admit multistationarity if there exists a choice of parameters for which the system \eqref{eq:maineq} has multiple fixed points. Building on the work of Banaji and Pantea, the capacity of a network to exhibit multistationarity under parameter-rich kinetics can be characterized in terms of CS-matrices.

\begin{theorem}[\cite{BaPa16}, \cite{VasHunt}]\label{thm:multistationarity}
A nondegenerate consistent network endowed with parameter-rich kinetics admits multistationarity if and only if there exist two $(|M|-n)\times(|M|-n)$ CS-matrices $S[\pmb{\kappa}_1]$ and $S[\pmb{\kappa}_2]$ such that
\begin{equation}\label{eq:multistationarity}
    \operatorname{det}S[\pmb{\kappa}_1] \operatorname{det}S[\pmb{\kappa}_2] < 0.
\end{equation}
\end{theorem}
\proof
Banaji and Pantea \cite[Theorem 3]{BaPa16} proved that a nondegenerate consistent network admits multistationarity if the Jacobian of the system restricted to a stoichiometric compatibility class can be made singular for a certain choice of parameters.  \cite[Corollary 5.7]{VasHunt} presents the straightforward translation of this condition in terms of Child-Selections as \eqref{eq:multistationarity}.
\endproof

\paragraph{Purely-imaginary eigenvalues: Periodic Orbits.}

Periodic oscillations play a central role in many reaction networks. In biochemistry, they regulate metabolic processes, circadian rhythms, and other essential biological functions \cite{Hess71}. The following result \cite{blokhuis_stoichiometric_2025} gives a sufficient condition in terms of CS-matrices for the system \eqref{eq:maineq} to admit nonstationary periodic solutions.

\begin{theorem}[\cite{blokhuis_stoichiometric_2025}]\label{thm:oscipos}
Let $\mathbf{\Gamma}$ be a consistent, nondegenerate network endowed with parameter-rich kinetics. Assume there exists a CS-matrix $S[\pmb{\kappa}]$ that is Hurwitz-stable and possesses a strict $k'\times k'$ principal submatrix $S[\pmb{\kappa}']$, which is an unstable-positive feedback. Then there exists a choice of parameters for which system \eqref{eq:maineq} admits nonstationary periodic solutions.
\end{theorem}
Minimal CS-matrices satisfying the assumption of the theorem are called \emph{Oscillatory cores of Class A}. Other sufficient conditions, based e.g. on unstable-negative feedback are also provided in \cite{blokhuis_stoichiometric_2025}, which we state here for consistency.

\begin{theorem}[\cite{blokhuis_stoichiometric_2025}]\label{thm:oscineg}
Let $\mathbf{\Gamma}$ be a consistent, nondegenerate network endowed with parameter-rich kinetics. Assume there exists an unstable-negative feedback $S[\pmb{\kappa}]$ which  possesses a Hurwitz-stable $(k-1)\times (k-1)$ principal submatrix. Then there exists a choice of parameters for which system \eqref{eq:maineq} admits nonstationary periodic solutions.
\end{theorem}

Minimal CS-matrices satisfying the assumption of the theorem are called \emph{Oscillatory cores of Class B}. The proof of both results relies on the theory of global Hopf bifurcation \cite{Fiedler85PhD}.

\section{The \textsc{BiRNe} module}\label{sec:birne}

\textsc{BiRNe} is an easy-to-use, open-source, Python-based module designed to identify stability changes of fixed points in reaction networks and to determine whether they are (i) of autocatalytic or non-autocatalytic nature, and (ii) whether they induce multistationarity or periodic oscillations.
The module targets reaction networks of medium size (around 15 species or reactions).
\textsc{BiRNe} is built on \textsc{SymPy}, a computer algebra system for symbolic mathematical computations that provides a wide variety of tools and solvers ranging from calculus to linear algebra \cite{Sympy17}.
\textsc{BiRNe} accepts a list of reactions as \textbf{input}. See Algorithm \ref{alg:Centralization} for the pseudocode. In plain terms, its workflow proceeds as follows:

\begin{enumerate}
\item \textbf{Build the stoichiometric matrices.}
The input list of reactions is translated into \emph{reactant} and \emph{product} matrices, $S^-$ and $S^+$, respectively.
The \emph{stoichiometric matrix} $S$ is then obtained as $S = S^+ - S^-$. See \eqref{eq:reactantmatrix}, \eqref{eq:productmatrix}, \eqref{eq:stoichmatrix} for a definition of such matrices.

\item \textbf{Check the network’s consistency.}
The existence of a strictly positive right kernel vector of $S$, and hence the consistency of the network \eqref{eq:consistent}, is determined by solving the linear programming problem using SciPy’s linear optimizer \cite{SciPy20}:
\begin{equation}
\begin{split}
\min & \quad v \\
\mathrm{s.t.} \qquad Sv &= 0 \\
v &> 0
\end{split}
\end{equation}
In certain modeling contexts, such as metabolic networks, the lack of consistency may simply reflect the omission of obvious production or degradation reactions, or the deliberate focus on a specific subnetwork structure. Moreover, results such as Thm.~\ref{thm:oscipos} naturally extend to any larger consistent network that contains the same motifs. For these reasons, if the network is not consistent, a warning is issued, but the algorithm continues its execution. 

\item \textbf{Generate the symbolic reactivity and Jacobian matrices.}
The symbolic reactivity matrix $R$ is generated from the reactant matrix and has entries $R_{jm}$ of the form
\begin{equation}
R_{jm} =
\begin{cases}
\verb|r(j,m)|, & \text{if } S^-_{mj}>0,\\
0, & \text{otherwise.}
\end{cases}
\end{equation}
Here, \verb|r(j,m)| is a \textsc{SymPy} symbol instantiated with the string representation `\verb|r(j,m)|’, allowing instantaneous identification of its position in $R$.
The symbolic Jacobian $G$ is then computed as
\begin{equation}
G = SR.
\end{equation}

\item \textbf{Compute and organize the characteristic polynomial of $G$.}
The characteristic polynomial of $G$ is computed using \textsc{SymPy}’s \textsc{charpoly} function, which implements the Samuelson–Berkowitz algorithm \cite{berkowitz84}.
Coefficients $c_k$ are extracted by size.
Relating to \eqref{eq:charpoly} and Lemma~\ref{lem:CSexp}, they correspond to
\begin{equation}
c_k = (-1)^k a_k = \sum_{\pmb{\kappa}} (-1)^k \operatorname{det} S[\pmb{\kappa}] \prod_{X_m \in \kappa} \verb|r(j,m)|,
\end{equation}
where each \verb|r(j,m)| satisfies \verb|r(j,m)| = $R_{J(X_m)m}$ for some CS bijection $J$.
In particular, CB-summands associated with $\pmb{\kappa}$ whose CS-matrix $S[\pmb{\kappa}]$ is Hurwitz-unstable and satisfies
\begin{equation}
\operatorname{sign} \operatorname{det} S[\pmb{\kappa}] = (-1)^{k-1}
\end{equation}
are readily identified in $c_k$ as those carrying negative coefficients.

Moreover, particular emphasis is given to \emph{minimal} negative CB-summands, that is, summands
\begin{equation} 
(-1)^k \operatorname{det} S[\pmb{\kappa}] \prod_{X_m \in \kappa} \verb|r(j,m)| < 0
\end{equation}
in $c_k$ such that in \emph{no} coefficient $c_{k'}$ with $k' < k$ there exists a negative summand
\begin{equation} 
(-1)^{k'} \operatorname{det} S[\pmb{\kappa}'] \prod_{X_m \in \kappa'} \verb|r(j,m)| < 0,
\end{equation}
where $\pmb{\kappa}'$ is a restriction of $\pmb{\kappa}$, i.e.
\begin{equation}\label{eq:descendant}
    \{\verb|r(j,m)| \;|\; X_m\in \kappa'\} \subset \{\verb|r(j,m)| \;|\; X_m\in \kappa\}.
\end{equation}
The CS-matrix associated with such minimal negative CB-summands identifies the \emph{unstable cores}, which are \emph{unstable positive feedbacks} (see Def.~\ref{def:unstable-positive}).  In parallel, a Hasse diagram \cite{baker_partial_1972} is constructed with the root vertex $c_0 = 1$. The direct descendants of the root vertex are the minimal negative CB-summands representing unstable-positive feedbacks. From each such vertex, identified by a CS-triple $\pmb{\kappa'}$, further descendants are identified in coefficients $c_k$, $k > k'$, for CB-summands associated to CS-triples $\pmb{\kappa}$ whenever the relation \eqref{eq:descendant} holds, that is, when $\pmb{\kappa'}$ is a restriction of $\pmb{\kappa}$.

\item \textbf{Check nondegeneracy of the network.}
According to Prop.~\ref{prop:nondegnet} and Lemma~\ref{lem:CSexp}, nondegeneracy is verified by ensuring that the largest $(|M| - \mathrm{dim}(\mathrm{Ker}\, S^T))^{\mathrm{th}}$ coefficient is not identically zero as a function of the symbolic entries of $R$. If the network is found to be degenerate, a warning is issued: most of the known results hold for nondegenerate networks. For analogous modeling reasons as exposed in the consistency check above, the algorithm continues nevertheless its execution.

\item \textbf{Check if the network admits stability.}
Likewise, the same largest $(|M| - \mathrm{dim}(\mathrm{Ker}\, S^T))^{\mathrm{th}}$ coefficient is examined to determine if the network admits stability.
For any CB-summand with positive sign, the corresponding CS-matrix is extracted and its eigenvalues computed.
If any such CS-matrix is Hurwitz-stable, the network is declared stable and further testing is omitted.

\item \textbf{Test for multistationarity.}
We check whether the same coefficient contains CB-summands with both positive and negative signs.
If this is the case, for nondegenerate networks, multistationarity is implied by Thm.~\ref{thm:multistationarity}.

\item \textbf{Testing autocatalysis.} 
Negative CB-summands are further examined for autocatalysis: the associated CS-matrix is extracted and tested for the Metzler property. 
If the matrix is Metzler, the corresponding feedback is autocatalytic. 
Autocatalytic unstable-positive feedbacks identify \emph{autocatalytic cores} in the sense of \cite{blokhuis20}. 
A network is certified as \emph{autocatalytic} if at least one negative CB-summand is associated with a Metzler CS-matrix, and as \emph{non-autocatalytic} otherwise.

\item \textbf{Testing oscillations.} 
For any \emph{negative} CB-summand in a coefficient $c_k$, 
\begin{equation} 
(-1)^k \operatorname{det} S[\pmb{\kappa}] \prod_{X_m \in \kappa} \verb|r(j,m)| < 0,
\end{equation}
we test whether there exists a CB-summand with \emph{positive} sign in a coefficient $c_{\tilde{k}}$ with $\tilde{k} > k$,
\begin{equation} 
(-1)^{\tilde{k}} \operatorname{det} S[\pmb{\tilde{\kappa}}] \prod_{X_m \in \tilde{\kappa}} \verb|r(j,m)| > 0,
\end{equation}
such that $\pmb{\kappa}$ is a restriction of $\tilde{\pmb{\kappa}}$, i.e.
\begin{equation}
    \{\verb|r(j,m)| \;|\; X_m\in \kappa\} \subset \{\verb|r(j,m)| \;|\; X_m\in \tilde{\kappa}\}.
\end{equation}
This corresponds to the existence of an edge in the constructed Hasse diagram that connects vertex with negative coefficient to a descendant vertex with positive coefficient. Once such a pair is found, the Hurwitz-stability of $S[\pmb{\tilde{\kappa}}]$ is checked. 
If stability is confirmed, $S[\pmb{\tilde{\kappa}}]$ satisfies the assumptions of Thm.~\ref{thm:oscipos}, and the network is certified to admit nonstationary periodic solutions.
\end{enumerate}

\begin{algorithm}[H]
    \small
   		\caption{\textsc{BiRNe}}
   		\label{alg:Centralization}
   		\begin{algorithmic}
 			\Require{$L$: List of reactions}
            \Ensure{$\mathcal{U}_{na}, \mathcal{U}_a, \mathcal{O}$: map of unstable (non-)autocatalytic cores and oscillatory  cores}
            \State $\mathcal{U}_a, \mathcal{U}_{n}, \mathcal{O} \leftarrow \langle \text{CB-summand}, \text{CS-matrix}\rangle, \langle \text{CB-summand}, \text{CS-matrix}\rangle, \langle \text{CB-summand}, [\ ]\rangle$
            \State $S^+, S-, M, R \leftarrow $\Call{StoichiometricMatrices}{$L$} \Comment{$M$: metabolites, $R$: reactions}
            \State $S \leftarrow S^+ - S^-$
            \State $\mathrm{admitsStability}, \ \mathrm{multiStationarity} \leftarrow \mathrm{False}, \mathrm{False}$
 			\If{$\exists v\in \mathbb{R}_{>0}: Sv=0$} \Comment{Consistency check}
                \State $m \leftarrow |M|$
                \State $d \leftarrow \mathrm{dim}(\mathrm{Ker} S^T)$
                \State $\mathbf{R} \leftarrow$ \Call{GenerateReactivityMatrix}{$S^-$}
     			\State $\mathbf{G} \leftarrow S\cdot \mathbf{R}$ 
     			\State $p \leftarrow $ \Call{Charpoly}{$\mathbf{G}$}	\Comment{$p=\mathbf{G}.\mathrm{charpoly}(\lambda)$}
     			\State $C \leftarrow $ \Call{Coefficients}{$p$}			\Comment{$C=\langle int, list \rangle$}
     			\If{$C[m-d] \neq \emptyset$}                            \Comment{Non-degeneracy check}
                    \State $C_n, C_p \leftarrow $ \Call{DivideCoefficientsBySign}{$C$} \Comment{$C_{n/p} = \langle int, list\rangle$}
                    \If{$C_n[m] \neq \emptyset \ \mathrm{and} \ C_p[m] \neq \emptyset$}\Comment{Multistationarity check}
                        \State $\mathrm{multistationarity} \ \leftarrow \mathrm{True}$
                    \EndIf    
                    \For{$c \in C_p[m]$}                                       \Comment{Stability check}
                        \State $S_c \leftarrow$ \Call{CSMatrix}{$S, c$}
                        \If{$S_c$ is Hurwitz-stable}
                            \State $\mathrm{admitsStability} \leftarrow \mathrm{True}$
                            \State Break
                        \EndIf
                    \EndFor
                    \For{$i \in \{1,\ldots, k-1\}$}
                          \For{$c \in C_n[i]$}
                            \If{\Call{MinimalityCheck}{$c$}$==$True}
                                \State $S_c \leftarrow $ \Call{CSMatrix}{$S, c$}
                                \State $k \leftarrow |S_c|$
                                \If{$S_c$ is Metzler}\Comment{Autocatalysis check}
                                    \State $\mathrm{autocatalytic} \leftarrow \mathrm{True}$ 
                                    \State $\mathcal{U}_a[c] \leftarrow S_c$
                                \Else
                                    \State $\mathcal{U}_{n}[c] \leftarrow S_c$
                                  \EndIf
                                \For {$\ell \in \{i+1, \ldots, m\}$}
                                    \For{$c' \in C_p[\ell]$}
                                        \State $S_{c'}\leftarrow $\Call{CSMatrix}{$S, c'$}
                                        \If{$c\subseteq c'$ and $S_{c'}$ is Hurwitz-stable}     \Comment{Oscillatory Check}
                                            \If{\Call{MinimalityCheck}{$c'$}
                                            $==$ True}
                                                \State $\mathcal{O}[c]\leftarrow [S_{c}, c', S_{c'}]$
                                            \EndIf
                                        \EndIf
                                    \EndFor
                                \EndFor
                            \EndIf
                        \EndFor
                    \EndFor
                \EndIf
            \EndIf
     		\State \Return $\mathcal{U}_a, \mathcal{U}_{n}, \mathcal{O}, \ \mathrm{admitsStability}$
 		\end{algorithmic}
 	\end{algorithm}

\section{An example}\label{sec:example}

To show the effectiveness of \textsc{BiRNe}, we study a simplified model of glycolysis and the pentose phosphate pathway in the central metabolism of \textit{E.~coli}. As the purpose of this example is here only to demonstrate the capabilities of our module, we do not discuss any biological implication. The network consists of 14 metabolites: Glucose, Glucose-6-Phosphate (G6P), 6-Phosphogluconolactone (6PG), Fructose-6-Phosphate (F6P), Fructose-1,6-Bi\-sphosphate (F16P), Dihydroxyacetonephosphate (DHAP), Glyceraldehyde-3-Phos\-phate (G3P), Xylulose-5-Phosphate (X5P), Ribose-5-Phosphate (R5P), Erythrose-4-Phos\-phate (E4P), Sedoheptulose-7-Phosphate (S7P), Sedoheptulose-1,7-Bisphosphate (S17P), Phosphoenolpyruvate (PEP), and Pyruvate (PYR); together with the following 23 reactions among them.

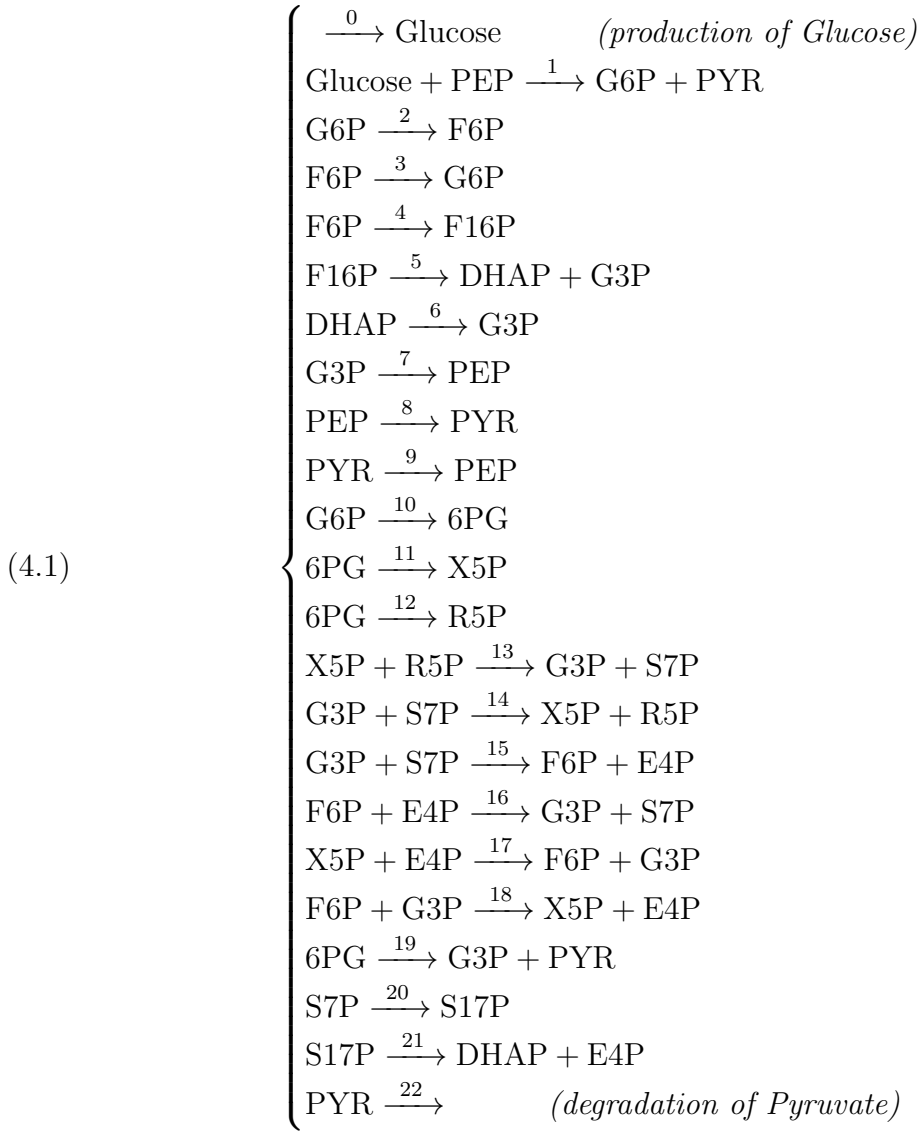
\begin{equation}
\begin{cases}
\;\ce{->[0]Glucose} \quad\quad\quad \textit{(production of Glucose)}\\
\ce{{Glucose} + {PEP} ->[1] {G6P} + {PYR}} \\
\ce{{G6P} ->[2] {F6P}} \\
\ce{{F6P} ->[3] {G6P}} \\
\ce{{F6P} ->[4] {F16P}} \\
\ce{{F16P} ->[5] {DHAP} + {G3P}} \\
\ce{{DHAP} ->[6] {G3P}} \\
\ce{{G3P} ->[7] {PEP}} \\
\ce{{PEP} ->[8] {PYR}} \\
\ce{{PYR} ->[9] {PEP}} \\
\ce{{G6P} ->[10] {6PG}} \\
\ce{{6PG} ->[11] {X5P}} \\
\ce{{6PG} ->[12] {R5P}} \\
\ce{{X5P} + {R5P} ->[13] {G3P} + {S7P}} \\
\ce{{G3P} + {S7P} ->[14] {X5P} + {R5P}} \\
\ce{{G3P} + {S7P} ->[15] {F6P} + {E4P}} \\
\ce{{F6P} + {E4P} ->[16] {G3P} + {S7P}} \\
\ce{{X5P} + {E4P} ->[17] {F6P} + {G3P}} \\
\ce{{F6P} + {G3P} ->[18] {X5P} + {E4P}} \\
\ce{{6PG} ->[19] {G3P} + {PYR}} \\
\ce{{S7P} ->[20] {S17P}} \\
\ce{{S17P} ->[21] {DHAP} + {E4P}}\\
\ce{PYR ->[22]} \quad\quad\quad \textit{(degradation of Pyruvate)}
\end{cases}
\end{equation}

The size and connectivity of this metabolic network seats at the upper end of the applicability of \textsc{BiRNe}. The module computed the characteristic polynomial in 18,651 s and constructed the Hasse diagram and determined all unstable cores in further 8,037 s. The network was found to be consistent, non-degenerate, and to admit both stability and multistationarity. In total, 33,820 negative CB-summands were identified, comprising 49 unstable cores: 29 autocatalytic and 20 non-autocatalytic. Among these unstable cores, 8 (5 autocatalytic and 3 non-autocatalytic) gave rise to 11 oscillatory motifs.

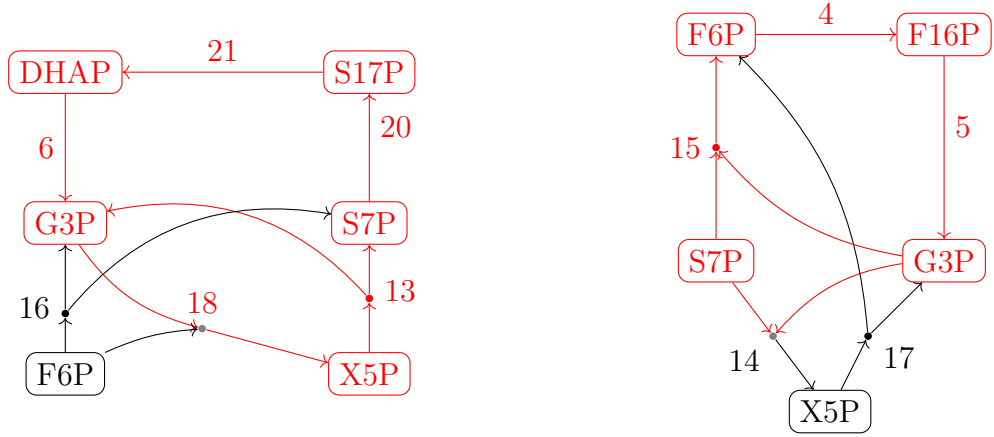
\begin{figure}
    \begin{minipage}[c]{0.6\textwidth}
        \centering
 	     \begin{tikzpicture}
    \node[draw, red, rectangle, rounded corners] (a) at (0,0) {DHAP};
    \node[draw, red, rectangle, rounded corners] (b) at (0,-2) {G3P};
    \node[draw, red, rectangle, rounded corners] (c) at (4,-4) {X5P};
    \node[draw, red, rectangle, rounded corners] (d) at (4,-2) {S7P};
    \node[draw, red, rectangle, rounded corners] (e) at (4,0) {S17P};
    \node[draw, rectangle, rounded corners] (f) at (0,-4) {F6P};

    \draw[->, red, rounded corners] (a) -- (b) node[midway, left] {6};
    \draw[->, red, rounded corners] (d) -- (e) node[midway, above right] {20};
    \draw[->, red, rounded corners] (e) -- (a) node[midway, above] {21};

    \node[circle, fill=red, inner sep=1pt] (h14) at (4,-3) {};
    \node[red, anchor=west, yshift=3pt] at (h14.east) {13};
    \draw[->, red, rounded corners] (c) -- (h14);
    \draw[->, red, rounded corners] (h14) -- (d);
    \draw[->, red, rounded corners] (h14) to [bend right=30] (b);

    \node[circle, fill=black, inner sep=1pt] (h15) at (0,-3.2) {};
    \node[anchor=east, yshift=2pt] at (h15.west) {16};
    \draw[->, rounded corners] (f) -- (h15);
    \draw[->, rounded corners] (h15) -- (b);
    \draw[->, rounded corners] (h15) to [bend left=30] (d);

    \node[circle, fill=gray, inner sep=1pt] (h16) at (1.8,-3.4) {};
    \node[red, anchor=south] at (h16.north) {18};
    \draw[->, red, rounded corners] (b) to [bend right=20] (h16);
    \draw[->, rounded corners] (f) to [bend left=10] (h16);
    \draw[->, red, rounded corners] (h16) -- (c);
\end{tikzpicture}
    \end{minipage}%
    \hfill%
 	\begin{minipage}[c]{0.4\textwidth}
        \centering
 		\begin{tikzpicture}
 			\node[draw, red, rectangle, rounded corners] (a) at (0,0) {F6P};
 			\node[draw, red, rectangle, rounded corners] (b) at (3,0) {F16P};
 			\node[draw, red, rectangle, rounded corners] (c) at (3,-3) {G3P};
 			\node[draw, red, rectangle, rounded corners] (d) at (0,-3) {S7P};
 			\node[draw, rectangle, rounded corners] (e) at (1.5,-5) {X5P};
            \node[circle, fill=black, inner sep=1pt, label=-10:17] (f) at (2.0,-4) {};
            \node[circle, fill=red, inner sep=1pt, label={[text=red]-180:15}] (g) at (0,-1.5) {};
            [circle, fill=gray, inner sep=1pt]
            \node[circle, fill=gray, inner sep=1pt, label=-120:14] (h) at (0.75,-4) {};
 			\draw[->, red, rounded corners] (a) -- (b) node[midway, above] {4};
 			\draw[->, red, rounded corners] (b) -- (c) node[midway, above right] {5};
 			\draw[->, red, rounded corners] (d) to (h);
 			\draw[->, red, rounded corners] (c) to [bend right=20] (h);
 			\draw[->, rounded corners] (e) to (f);
 			\draw[->, rounded corners] (f) to (c);
            \draw[->, rounded corners] (h) to (e);
            \draw[->, rounded corners] (f) to [bend right = 20](a);
            \draw[->, red, rounded corners] (d) to (g);
            \draw[->, red, rounded corners] (c) to [bend left = 20](g);
            \draw[->, red, rounded corners] (g) to (a);
            
 		\end{tikzpicture}
 	\end{minipage}

    \caption{Depiction of two oscillatory cores of class A based on an autocatalytic (left) and a non-autocatalytic (right) unstable-positive feedback within the Glycolysis and Pentose Phosphate Pathway. Both unstable-positive feedbacks are shown in red. For clarity, hyperarrows involving multiple reactants or products are represented here in bipartite form. The associated CS-matrices are, respectively, \eqref{eq:Sex1} and \eqref{eq:Sex2}.} 
    \label{fig:Oscillatory}
\end{figure}

We refer to \url{https://github.com/hollyritch/Bi.R.Ne} for the complete output. Here we present two oscillatory cores of class A to exemplify: one based on autocatalytic unstable-positive feedback and one on non-autocatalytic unstable-positive feedback (see Figure~\ref{fig:Oscillatory}, left and right, respectively). The first example, left in Fig.~\ref{fig:Oscillatory},  concerns the following CS-triple:
\begin{equation}
\begin{split}
\pmb{\kappa}_1=(&\kappa_1=\{\ce{{F6P}},\ce{DHAP},\ce{{G3P}},\ce{{X5P}},\ce{{S7P}},\ce{{S17P}}\},\;E_{\kappa_1}=\{6,13,16,18,20,21\},\\
&J(\kappa)=\{16,6,18,13,20,21\}),
\end{split}
\end{equation}
with associated Hurwitz-stable CS-matrix:
\begin{equation}\label{eq:Sex1}
S[\pmb{\kappa}_1]=
\begin{blockarray}{ccccccc}
& 16 & 6 & 18 & 13 & 20 & 21\\
\begin{block}{c(cccccc)}
\ce{{F6P}} & -1 & 0 & -1 & 0 & 0 & 0\\
\ce{{DHAP}}& 0 & -1 & 0 & 0 & 0 & 1\\
\ce{{G3P}} & 1 & 1 & -1 & 1 & 0 & 0\\
\ce{{X5P}} & 0 & 0 & 1 & -1 & 0 & 0\\
\ce{{S7P}} & 1 & 0 & 0 & 1 & -1 & 0\\
\ce{{S17P}}& 0 & 0 & 0 & 0 & 1 & -1\\
\end{block}
\end{blockarray},
\end{equation}
with eigenvalues all $\lambda_i=-1$. On the other hand, marked in red in Fig.~\ref{fig:Oscillatory}, the restriction $\pmb{\kappa}'_1$ defined on $\kappa_1 \setminus \{\ce{{F6P}}\}$:
\begin{equation}
\begin{split}
\pmb{\kappa}'_1=(&\kappa'_1=\{\ce{DHAP},\ce{{G3P}},\ce{{X5P}},\ce{{S7P}},\ce{{S17P}}\},\;E_{\kappa_1}=\{6,13,18,20,21\},\\
&J(\kappa_1)=\{6,18,13,20,21\}),
\end{split}
\end{equation}
identifies a Hurwitz-unstable CS-matrix, which is in particular an unstable-positive feedback and Metzler. This yields autocatalysis via Thm.~\ref{thm:autocat}. 
\begin{equation}
S[\pmb{\kappa}'_1]=
\begin{blockarray}{cccccc}
& 6 & 18 & 13 & 20 & 21\\
\begin{block}{c(ccccc)}
\ce{{DHAP}} & -1 & 0 & 0 & 0 & 1\\
\ce{{G3P}} & 1 & -1 & 1 & 0 & 0\\
\ce{{X5P}} & 0 & 1 & -1 & 0 & 0\\
\ce{{S7P}} & 0 & 0 & 1 & -1 & 0\\
\ce{{S17P}}& 0 & 0 & 0 & 1 & -1\\
\end{block}
\end{blockarray},
\end{equation}
with $\operatorname{det}S[\pmb{\kappa}'_1]=(-1)^{5-1}=1$. The CS-matrix $S[\pmb{\kappa}'_1]$ corresponds to an autocatalytic core sensu \cite{blokhuis20}, precisely of type II. In conclusion, $S[\pmb{\kappa}_1]$ is an oscillatory core of class A.

The second example, right in Fig.~\ref{fig:Oscillatory}, is based on the following CS-triple:
\begin{equation}
\begin{split}
\pmb{\kappa}_2=(&\kappa_2=\{\ce{{F6P}},\ce{{F16P}},\ce{{G3P}},\ce{{X5P}},\ce{{S7P}}\},\;E_{\kappa_2}=\{4,5,14,15,17\},\\
&J(\kappa_2)=\{4,5,15,17,14\})
\end{split}
\end{equation}
with associated Hurwitz-stable CS-matrix:
\begin{equation}\label{eq:Sex2}
S[\pmb{\kappa}_2]=
\begin{blockarray}{cccccc}
 & 4 & 5 & 15 & 17 & 14\\
\begin{block}{c(ccccc)}
\ce{{F6P}} & -1 & 0 & 1 & 1 & 0 \\
\ce{{F16P}}& 1 & -1 & 0 & 0 & 0 \\
\ce{{G3P}} & 0 & 1 & -1 & 1 & -1 \\
\ce{{X5P}} & 0 & 0 & 0 & -1 & 1 \\
\ce{{S7P}} & 0 & 0 & -1 & 0 & -1\\
\end{block}
\end{blockarray},
\end{equation}
with eigenvalues $(\lambda_1,\lambda_2,\lambda_3,\lambda_4,\lambda_5)\approx (-2.24,-1.34\pm0.8i,-0.04\pm 0.43)$. On the other hand, marked in red in Fig.~\ref{fig:Oscillatory}, the restriction $\pmb{\kappa}'_2$ defined on $\kappa_2 \setminus \{\ce{{X5P}}\}$:
\begin{equation}
\begin{split}
\pmb{\kappa}'_2=(&\kappa'_2=\{\ce{{F6P}},\ce{{F16P}},\ce{{G3P}},\ce{{S7P}}\},\;E_{\kappa_2}=\{4,5,14,15\},\\
&J(\kappa_2)=\{4,5,15,14\})
\end{split}
\end{equation}
identifies a Hurwitz-unstable CS-matrix $S[\pmb{\kappa}'_2]$, which is in particular an unstable-positive feedback but not a Metzler matrix; thus, there is no autocatalysis:
\begin{equation}
S[\pmb{\kappa}'_2]=
\begin{blockarray}{ccccc}
 & 4 & 5 & 15 & 14\\
\begin{block}{c(cccc)}
\ce{{F6P}} & -1 & 0 & 1 & 0 \\
\ce{{F16P}}& 1 & -1 & 0 & 0 \\
\ce{{G3P}} & 0 & 1 & -1 & -1 \\
\ce{{S7P}} & 0 & 0 & -1 & -1\\
\end{block}
\end{blockarray},
\end{equation}
with $\operatorname{det}S[\pmb{\kappa}'_2]=(-1)^{4-1}=-1$. Finally, we point out that the same set $\kappa_2$ and $E_\kappa$, but with different bijection $J(\kappa_2)=\{4,5,17,15,14\}$, gives rise to a Hurwitz-unstable CS-matrix:
\begin{equation}
S[\pmb{\tilde{\kappa}}_2]=
\begin{blockarray}{cccccc}
 & 4 & 5 & 14 & 17 & 15\\
\begin{block}{c(ccccc)}
\ce{{F6P}} & -1 & 0 & 0 & 1 & 1 \\
\ce{{F16P}}& 1 & -1 & 0 & 0 & 0 \\
\ce{{G3P}} & 0 & 1 & -1 & 1 & -1 \\
\ce{{X5P}} & 0 & 0 & 1 & -1 & 0 \\
\ce{{S7P}} & 0 & 0 & -1 & 0 & -1\\
\end{block}
\end{blockarray},
\end{equation}
obtainable from $S[\pmb{\kappa}_2]$ with a single swap of columns (14,15). In particular, $S[\pmb{\tilde{\kappa}}_2]$ is unstable and does not give rise to oscillations. For simplicity, we have not marked such subtlety in Fig.~\ref{fig:Oscillatory}.

\section{Discussion}\label{sec:outlook}

We have presented \textsc{BiRNe}, a Python-based computer algebra module designed to identify motifs associated with stability changes of fixed points in reaction networks and the resulting bifurcations, classified as either leading to multistationarity (zero-eigenvalue bifurcations) or to periodic oscillations (purely imaginary eigenvalue bifurcations).

In contrast to commercial symbolic computing environments with restrictive licensing and recurring fees, Python is developed under an OSI-approved open-source license, making it freely accessible. In addition, \textsc{SymPy} provides a powerful computer algebra system within Python for symbolic computations, offering a broad range of algorithms and solvers for calculus and linear algebra. For networks of moderate size (typically up to 15 species and reactions, depending on the network’s connectivity) our implementation can determine multistationarity and identify all unstable-positive feedbacks, classifying them as autocatalytic or non-autocatalytic, and further detecting whether they give rise to periodic oscillations.  The presented algorithm relies solely on basic functionality common to most symbolic computation platforms. Consequently, it could be adapted to more efficient symbolic programming environments with minimal effort. On the other hand, substantial performance gains, such as those needed to apply the method to large-scale networks, appear unlikely: the approach relies on computing the characteristic polynomial of a symbolic matrix, an operation whose computational complexity grows exponentially with the matrix size, i.e., with the number of species in the network. Thus, at best, only minor gains would be expected, i.e. size limitations could be raised only by a handful of additional species and reactions within reasonable computation times. The analysis of large reaction networks, such as metabolic systems in bacteria or eukaryotes involving hundreds to thousands of species and reactions, requires a fundamentally different methodological approach. In this regard, we refer to our forthcoming framework for detecting autocatalytic cycles in chemical reaction networks \cite{Golnik25}, based on graph-theory tools.

The current version (November 2025) of the module evaluates multistationarity based on Thm.~\ref{thm:multistationarity} without specifying the bifurcation type and scans only for one mechanism of oscillations (Thm.~\ref{thm:oscipos}). Future developments aim to extend these capabilities by certifying specific zero-eigenvalue bifurcations, such as \emph{saddle-node bifurcations} (cf.~\cite{V23}), and incorporating additional oscillatory motifs, for example those described as Recipe 0 in \cite{blokhuis_stoichiometric_2025} or addressed in Thm.~\ref{thm:oscineg}, thereby broadening the module’s analytical scope and applicability.

\bibliography{references.bib}

@article{Fiedler85PhD,
  title={An index for global {H}opf bifurcation in parabolic systems.},
  author={Fiedler, Bernold},
  journal={Journal f{\"u}r die reine und angewandte Mathematik},
  volume={358},
  pages={1--36},
  year={1985}
}

@article{MA64,
  title={Studier over affiniteten},
  author={Waage, P and Guldberg, CM},
  journal={Forhandlinger i Videnskabs-selskabet i Christiania},
  volume={1},
  pages={35--45},
  year={1864}
}

@article{MESSI:18,
  author =       {P{\'e}rez Mill{\'a}n, Mercedes and Dickenstein, Alicia},
  title =        {The Structure of {MESSI} Biological Systems},
  journal =      {SIAM Journal on Applied Dynamical Systems},
  volume =       {17},
  number =       {2},
  pages =        {1650-1682},
  year =         {2018},
  doi =          {10.1137/17M1113722}
}

@Article{Muller:12,
  author =       {M{\"u}ller, Stefan and Regensburger, Georg},
  title =        {Generalized mass action systems: Complex balancing
                  equilibria and sign vectors of the stoichiometric
                  and kinetic-order subspaces},
  journal =      {SIAM Journal on Applied Mathematics},
  volume =       {72},
  number =       {6},
  pages =        {1926-1947},
  year =         {2012},
  doi =          {10.1137/1108470},
}

@article{blokhuis20,
  title =        {Universal motifs and the diversity of autocatalytic systems},
  author =       {Blokhuis, Alex and Lacoste, David and Nghe, Philippe},
  journal =      {Proceedings of the National Academy of Sciences},
  volume =       {117},
  number =       {41},
  pages =        {25230-25236},
  year =         {2020},
  doi =          {10.1073/pnas.201352711},
}

@book{Bullo20,
  title =        {Lectures on network systems},
  author =       {Bullo, Francesco},
  volume =       {1},
  year =         {2020},
  publisher =    {Kindle Direct Publishing}
}

@book{Hsubook,
  title =        {Ordinary differential equations with applications},
  author =       {Hsu, Sze-Bi and Chen, Kuo-Chang},
  volume =       {23},
  year =         {2022},
  publisher =    {World scientific}
}

@article{Holling65,
  title={The functional response of predators to prey density and its role in mimicry and population regulation},
  author={Holling, Crawford Stanley},
  journal={The Memoirs of the Entomological Society of Canada},
  volume={97},
  number={S45},
  pages={5--60},
  year={1965},
  publisher={Cambridge University Press}
}

@article{BaBo23,
  title={The smallest bimolecular mass action reaction networks admitting {A}ndronov--{H}opf bifurcation},
  author={Banaji, Murad and Boros, Bal{\'a}zs},
  journal={Nonlinearity},
  volume={36},
  number={2},
  pages={1398},
  year={2023},
  publisher={IOP Publishing}
}

@article{VasStad23,
  title={Unstable cores are the source of instability in chemical reaction networks},
  author={Vassena, Nicola and Stadler, Peter F},
  journal={Proceedings of the Royal Society A},
  volume={480},
  number={2285},
  pages={20230694},
  year={2024},
  publisher={The Royal Society}
}

@article{V23,
author = {Vassena, Nicola},
title = {Structural Conditions for Saddle-Node Bifurcations in Chemical Reaction Networks},
journal = {SIAM Journal on Applied Dynamical Systems},
volume = {22},
number = {3},
pages = {1639-1672},
year = {2023},
doi = {10.1137/22M1527933}}

@article{VasHunt,
 author =       {Vassena, Nicola},
  title =       {Symbolic hunt of instabilities and bifurcations in
                  reaction networks},
  journal =     {Discrete and Continuous Dynamical Systems - Series B},
  doi =         {10.3934/dcdsb.2023190},
  year =        {2023},
  volume =      {30},
  number =      {6},
  pages =       {2183-2208},
}

@article{Hill10,
  title={The possible effects of the aggregation of the molecules of haemoglobin on its dissociation curves},
  author={Hill, Archibald Vivian},
  journal={The Journal of Physiology},
  volume={40},
  pages={4--7},
  year={1910}
}

@article{Ang07,
  title={A {P}etri net approach to the study of persistence in chemical reaction networks},
  author={Angeli, David and De Leenheer, Patrick and Sontag, Eduardo D},
  journal={Mathematical biosciences},
  volume={210},
  number={2},
  pages={598--618},
  year={2007},
  publisher={Elsevier}
}

@article{MM13,
  title={Die {K}inetik der {I}nvertinwirkung},
  author={Michaelis, L. and Menten, M. L.},
  journal={Biochem. Z.},
  volume={49},
  pages={333-369},
  year={1913},
}

@inproceedings{Conradi2007,
  title={Saddle-node bifurcations in biochemical reaction networks with mass action kinetics and application to a double-phosphorylation mechanism},
  author={Conradi, Carsten and Flockerzi, Dietrich and Raisch, Jorg},
  booktitle={2007 American control conference},
  pages={6103--6109},
  year={2007},
  organization={IEEE}
}

@article{Hess71,
  title={Oscillatory phenomena in biochemistry},
  author={Hess, Benno and Boiteux, Arnold},
  journal={Annual review of biochemistry},
  volume={40},
  number={1},
  pages={237--258},
  year={1971},
  publisher={Annual Reviews 4139 El Camino Way, PO Box 10139, Palo Alto, CA 94303-0139, USA}
}

@article{ThomKauf01,
  title={Multistationarity, the basis of cell differentiation and memory. {I}. {S}tructural conditions of multistationarity and other nontrivial behavior},
  author={Thomas, Ren{\'e} and Kaufman, Marcelle},
  journal={Chaos: An Interdisciplinary Journal of Nonlinear Science},
  volume={11},
  number={1},
  pages={170--179},
  year={2001},
  publisher={American Institute of Physics}
}

@article{BaPa16,
  title={Some results on injectivity and multistationarity in chemical reaction networks},
  author={Banaji, Murad and Pantea, Casian},
  journal={SIAM Journal on Applied Dynamical Systems},
  volume={15},
  number={2},
  pages={807--869},
  year={2016},
  publisher={SIAM}
}

@article{Gat2005,
  title={Toric ideals and graph theory to analyze {H}opf bifurcations in mass action systems},
  author={Gatermann, Karin and Eiswirth, Markus and Sensse, Anke},
  journal={Journal of Symbolic Computation},
  volume={40},
  number={6},
  pages={1361--1382},
  year={2005},
  publisher={Elsevier}
}

@article{rost21,
  title={Exotic bifurcations in three connected populations with Allee effect},
  author={R{\"o}st, Gergely and Sadeghimanesh, AmirHosein},
  journal={International Journal of Bifurcation and Chaos},
  volume={31},
  number={13},
  pages={2150202},
  year={2021},
  publisher={World Scientific}
}

@article{Golnik25,
    author = {Golnik, Richard and Gatter, Thomas and Stadler, Peter F and Vassena, Nicola},
    title = {Detecting autocatalytic cycles in chemical reaction networks},
    journal = {In preparation}
}

@article{hernandez23,
  title={A framework for deriving analytic steady states of biochemical reaction networks},
  author={Hernandez, Bryan S and Lubenia, Patrick Vincent N and Johnston, Matthew D and Kim, Jae Kyoung},
  journal={PLoS computational biology},
  volume={19},
  number={4},
  pages={e1011039},
  year={2023},
  publisher={Public Library of Science San Francisco, CA USA}
}

@misc{blokhuis_stoichiometric_2025,
  title = {Stoichiometric Recipes for Periodic Oscillations in Reaction Networks},
  author = {Blokhuis, Alex and Stadler, Peter F. and Vassena, Nicola},
  year = {2025},
  month = aug,
  number = {arXiv:2508.15273},
  eprint = {2508.15273},
  primaryclass = {q-bio},
  publisher = {arXiv},
  doi = {10.48550/arXiv.2508.15273},
  urldate = {2025-10-01},
  archiveprefix = {arXiv}
}

@article{Sympy17,
     title = {SymPy: symbolic computing in Python},
     author = {Meurer, Aaron and Smith, Christopher P. and Paprocki, Mateusz and \v{C}ert\'{i}k, Ond\v{r}ej and Kirpichev, Sergey B. and Rocklin, Matthew and Kumar, AMiT and Ivanov, Sergiu and Moore, Jason K. and Singh, Sartaj and Rathnayake, Thilina and Vig, Sean and Granger, Brian E. and Muller, Richard P. and Bonazzi, Francesco and Gupta, Harsh and Vats, Shivam and Johansson, Fredrik and Pedregosa, Fabian and Curry, Matthew J. and Terrel, Andy R. and Rou\v{c}ka, \v{S}t\v{e}p\'{a}n and Saboo, Ashutosh and Fernando, Isuru and Kulal, Sumith and Cimrman, Robert and Scopatz, Anthony},
     year = 2017,
     month = jan,
     volume = 3,
     pages = {e103},
     journal = {PeerJ Computer Science},
     issn = {2376-5992},
     url = {https://doi.org/10.7717/peerj-cs.103},
     doi = {10.7717/peerj-cs.103}
    }

@ARTICLE{SciPy20,
  author  = {Virtanen, Pauli and Gommers, Ralf and Oliphant, Travis E. and
            Haberland, Matt and Reddy, Tyler and Cournapeau, David and
            Burovski, Evgeni and Peterson, Pearu and Weckesser, Warren and
            Bright, Jonathan and {van der Walt}, St{\'e}fan J. and
            Brett, Matthew and Wilson, Joshua and Millman, K. Jarrod and
            Mayorov, Nikolay and Nelson, Andrew R. J. and Jones, Eric and
            Kern, Robert and Larson, Eric and Carey, C J and
            Polat, {\.I}lhan and Feng, Yu and Moore, Eric W. and
            {VanderPlas}, Jake and Laxalde, Denis and Perktold, Josef and
            Cimrman, Robert and Henriksen, Ian and Quintero, E. A. and
            Harris, Charles R. and Archibald, Anne M. and
            Ribeiro, Ant{\^o}nio H. and Pedregosa, Fabian and
            {van Mulbregt}, Paul and {SciPy 1.0 Contributors}},
  title   = {{{SciPy} 1.0: Fundamental Algorithms for Scientific
            Computing in Python}},
  journal = {Nature Methods},
  year    = {2020},
  volume  = {17},
  pages   = {261--272},
  adsurl  = {https://rdcu.be/b08Wh},
  doi     = {10.1038/s41592-019-0686-2},
}

@article{berkowitz84,
title = {On computing the determinant in small parallel time using a small number of processors},
journal = {Information Processing Letters},
volume = {18},
number = {3},
pages = {147-150},
year = {1984},
issn = {0020-0190},
doi = {https://doi.org/10.1016/0020-0190(84)90018-8},
url = {https://www.sciencedirect.com/science/article/pii/0020019084900188},
author = {Stuart J. Berkowitz},
}

@article{baker_partial_1972,
  title = {Partial Orders of Dimension 2},
  author = {Baker, K. A. and Fishburn, P. C. and Roberts, F. S.},
  year = {1972},
  month = jan,
  journal = {Networks},
  volume = {2},
  number = {1},
  pages = {11--28},
  issn = {0028-3045, 1097-0037},
  doi = {10.1002/net.3230020103},
  urldate = {2025-10-14},
  copyright = {http://onlinelibrary.wiley.com/termsAndConditions\#vor},
  langid = {english}
}
\bibliographystyle{amsplain}

\end{document}